# Gravitational Orbits and the Lambert Problem

Ian R. Gatland

Georgia Institute of Technology

October 15, 2021

ABSTRACT

The Lambert problem is to determine the gravitational orbit between two points that has a specified time of flight, allowing the second point to be a moving target such as a satellite. After a review of gravitational orbits, a solution of the Lambert problem is presented using the path equation directly. However, because the time interval is specified, the solution requires a search procedure. An example is given to illustrate the method.

INTRODUCTION

During the 18th century, Lambert introduced the problem of finding an orbital trajectory that satisfied Newton's Law of Gravity and went from one point to another point in a specified amount of time. A modern application could be a rocket carrying supplies to a space station. The starting point occurs when the rocket is above the atmosphere and the rocket motor has stopped. The end point is the rendezvous with the space station, and, since the space station is in orbit, the time of flight of the rocket must be known in order for the rendezvous to occur. Another application could be fixing the splashdown point for a rocket test. The target point is on the surface of the Earth (ocean) and this point moves due to the rotation of the Earth. The determination of the orbit converts a boundary value problem into an initial value problem. In particular, the solution determines the initial velocity, a significant feature in these examples and other practical applications.

One approach to this problem uses Lambert's Equation.[1] Several web pages discuss this in detail but require a comprehensive understanding of conic sections. Another, in the text by Bate et. al.,[2] uses Universal Variables. The solution method presented here uses the path equation and the relation between time and orbit angle directly.

GRAVITATIONAL ORBITS

Gravitational orbits are determined by Newton's Law of Gravity. For a single centrally-symmetric gravitational source the equation of motion for an orbiting body is

$$\vec{a} = -\frac{GM\vec{r}}{r^3}, \qquad (1)$$

where $\vec{a}$ is the acceleration of a body, $G$ is the gravitational constant, $M$ is the source mass, and $\vec{r}$ is the position of the orbiting body relative to the source center.

Equation (1) is an approximation. For a planet orbiting the Sun, it ignores the masses of the other planets. For an Earth satellite, it ignores the oblateness of the Earth. However, the solution of Eq. (1) is a valuable starting point for a more detailed analysis.[3] We can also approximate that the gravitational source center is an inertial reference frame if the orbiting body mass is negligible.

The first result from Eq. (1) is conservation of angular momentum, implying that the orbit is in a fixed plane and a two-dimensional analysis may be used.

The second result from Eq. (1), using conservation of energy or the Laplace-Runge-Lenz vector,[4] is the path equation

$$r = \frac{p}{(1 + \varepsilon \cos\theta)}, \qquad (2)$$

where $r$ is the radial distance, $\theta$ is the orbit angle, and $p$ and $\varepsilon$ are constants of integration. It is convenient, but not necessary, to recognize Eq. (2) as that of a conic section and identify $p$ as the perpendicular radius (semilatus rectum) and $\varepsilon$ as the eccentricity. The point of minimum radius, periapsis, is at $\theta = 0$.

The conservation of angular momentum and Eq. (2) lead to the relation between orbit angle and time, $t$, from periapsis:

$$\sqrt{\frac{GM}{p^3}}\, t = \int \frac{d\theta}{(1 + \varepsilon \cos\theta)^2}. \qquad (3)$$

This integral is evaluated using the half-angle expression for $\cos(\theta)$ in terms of $\tan(\theta/2)$ and a change of variable to an angle $\psi$ such that

$$\tan\left(\frac{\psi}{2}\right) = \sqrt{\frac{1-\varepsilon}{1+\varepsilon}} \tan\left(\frac{\theta}{2}\right). \qquad (4)$$

Then the solution is

$$\sqrt{\frac{GM}{p^3}}t = \frac{(\psi - \varepsilon \sin \psi)}{(1-\varepsilon^2)^{3/2}}. \tag{5}$$

Equations (4) and (5) apply for $\varepsilon < 1$. Similar equations for $\varepsilon = 1$ and $\varepsilon > 1$ may be obtained from Eq. (3). Equations (4) and (5) give the time directly from the orbit angle but obtaining the orbit angle from the time requires a search procedure to obtain $\psi$.

LAMBERT PROBLEM

The solution of the Lambert problem in the orbit plane has as specified quantities the initial and final radii $r_1$ and $r_2$, the angle $\gamma$ between $\vec{r}_1$ and $\vec{r}_2$, and the time-of-flight $t_f$. Figure 1 shows a possible elliptic orbit with a focus at the Earth center and going through the two points. Also shown are the radii for the two points, the bisector of the angle $\gamma$, and the major axis of the to-be-determined ellipse. The angle between the bisector and the major axis (toward apoapsis) is designated $\eta$. Also let $\chi = \pi - \gamma/2$. Then the angle of $\vec{r}_1$ from periapsis is $\chi - \eta$ and the angle of $\vec{r}_2$ back to periapsis is $\chi + \eta$.

Because the time-of-flight is specified, the solution of the Lambert problem involves a search on some parameter. Here $p$ is chosen as the search parameter. Applying Eq. (2) for the point $r_1$ where $\theta_1 = \chi - \eta$

$$\frac{p}{r_1} = 1 + \varepsilon \cos(\chi - \eta) = 1 + \varepsilon \cos \chi \cos \eta + \varepsilon \sin \chi \sin \eta \tag{6}$$

and for the point $r_2$ where $\theta_2 = 2\pi - (\chi + \eta)$

$$\frac{p}{r_2} = 1 + \varepsilon \cos(2\pi - [\chi + \eta]) = 1 + \varepsilon \cos \chi \cos \eta - \varepsilon \sin \chi \sin \eta \tag{7}$$

The sum of Eqs. (6) and (7) gives

$$\rho \equiv \frac{(p/r_1 + p/r_2 - 2)}{2\cos \chi} = \varepsilon \cos \eta \tag{8}$$

and the difference of Eqs. (6) and (7) gives

$$\sigma \equiv \frac{(p/r_1 - p/r_2)}{2\sin\chi} = \varepsilon \sin\eta \tag{9}$$

Equations (8) and (9) then yield

$$\varepsilon = \sqrt{\rho^2 + \sigma^2} \tag{10}$$

and

$$\eta = \arctan\left(\frac{\sigma}{\rho}\right) \tag{11}$$

with due regard to the signs of $\rho$ and $\sigma$. Now $p$, $\varepsilon$, $\chi$, and $\eta$ are known so $\theta_1 = \chi - \eta$ and $\theta_2 = 2\pi - (\chi + \eta)$. Hence the time $t_1$ from periapsis to $\vec{r}_1$ and the time $t_2$ from periapsis to $\vec{r}_2$ may be calculated using Eqs. (4) and (5). The time of flight between $\vec{r}_1$ and $\vec{r}_2$ is then $t = t_2 - t_1$. Finally, it is a case of varying $p$ to find $t = t_f$ by some suitable search procedure. The example below uses a simple method.

Applications often require the initial or final velocity. The velocity at an orbit angle $\theta$ has

$$v_r = \sqrt{\frac{GM}{p}}\,\varepsilon \sin\theta \tag{12}$$

as the radial component and

$$v_\theta = \sqrt{\frac{GM}{p}}(1 + \varepsilon \cos\theta) \tag{13}$$

as the transverse component.

EXAMPLE

Suppose $r_1$ = 6800 km, $r_2$ = 6400 km, $\gamma$ = 75°, and $t_f$ = 3000 s. Also, for the Earth, GM = 398600 (km)$^3$s$^{-2}$. Note $\chi$ = 2.4871 rad and all calculations are carried out using radian measurement of angles.

(1) For $p$ = 2500 km, Eqs. (8), (9), (10), (11), (4), and (5), in order, give $t$ = 3820 s.
(2) For $p$ = 3000 km, using the same procedure, $t$ = 2720 s.
(3) Linear interpolation between these two values for $t_f$ = 3000 s gives $p$ = 2873 km that has an actual time $t$ = 2925 s.

(4) Next interpolation between $p = 2500$ km and $p = 2873$ km provides $p = 2842$ km with $t = 2980$ s.

(5) Further investigation leads to $p = 2831.48$ km and a time of $t = 2999.99$ s.

For this value of $p$, $\varepsilon = 0.7195$ and $\eta = -0.0297$ rad ($-1.7°$). The initial velocity has a radial component, Eq. (12), 4.9936 km/s and a transverse component, Eq. (13), 4.9404 km/s.

Another example with $r_1 = 6800$ km, $r_2 = 6400$ km, $\gamma = 285°$, and $t_f = 6000$ s yields $p = 7589.79$ km and $\varepsilon = 0.1988$ for this long-way-round. The radial velocity component is 1.1692 km/s and the transverse component is 8.0886 km/s.

CONCLUSION

Gravitational orbits are determined by Newton's Law of Gravity. The equation of motion, Eq. (1), is all that is needed to obtain Eqs. (2) through (5).

Equations (6) to (11) provide the mathematical analysis to solve the Lambert problem. Equations (8) to (11) may be implemented in a computer program. Note that Eqs. (8) to (11) and Eqs. (4) and (5) are all single valued so the direct solution of the Lambert problem is unique.

The values of $p$ and $\varepsilon$ give the size and shape of the orbit so that it may be visualized. In particular, the periapsis radius is $p/(1+\varepsilon)$. As shown in the example, each step converts $(p, r_1, r_2, \gamma)$ to $(p, \varepsilon, \eta, t)$. Then the requirement that $t = t_f$ fixes the values of $p$, $\varepsilon$, and $\eta$.

The choice of $p$ as the search parameter is appropriate because it is a familiar physical quantity. A table of $t$, $p$, and $\varepsilon$ is valuable when selecting an initial trial value for $p$. If $p$, $r_1$, and $r_2$ are scaled by a factor $k$ then $\varepsilon$ and $\eta$ are unchanged and $t$ is scaled by a factor $k^{3/2}$, extending the value of the table.

According to Eq. (9), the angle $\eta$ between the bisector and the direction opposite periapsis (to apoapsis if it exists) is zero when $r_1 = r_2$ and, with Eq. (8), is small in many cases of interest. However, if a very short time-of-flight is specified the speed will exceed escape speed and the orbit will be hyperbolic.

Any presentation of gravitational orbits becomes more real with the addition of applications. One often used is the Hohmann transfer. The direct solution of the Lambert problem serves the same purpose and adds meaning to the phrase "launch window."

FIGURES

Fig. 1: A possible elliptic orbit passing through two points.

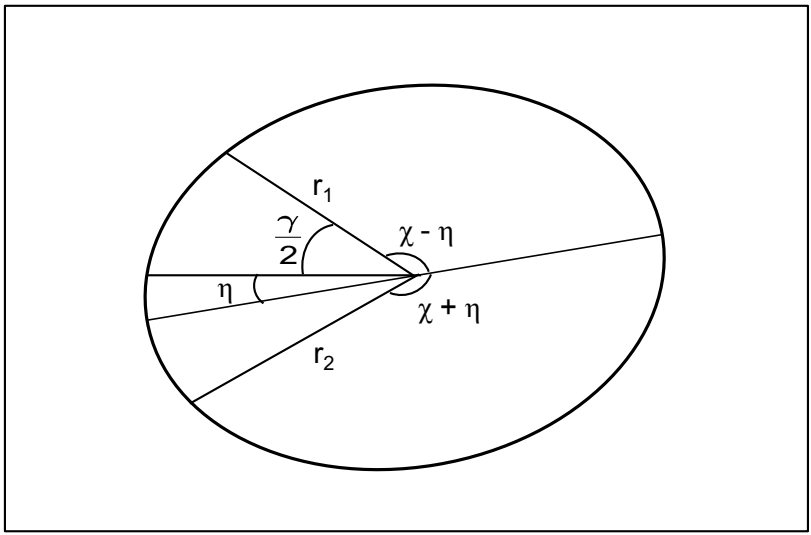

**Figure 1**